\documentclass[aps,prl,reprint,amsmath,superscriptaddress,longbibliography]{revtex4-1}
\usepackage[pdftex,bookmarks,colorlinks]{hyperref}
\pdfoutput=1
\usepackage[pdftex]{graphicx}
\usepackage{blindtext}
\usepackage{natbib}
\usepackage{upgreek}
\begin{document}

\title{Drop Shaping by Laser-Pulse Impact}

\author{Alexander L. Klein} \email[]{alexludwig.klein@utwente.nl}
\author{Wilco Bouwhuis}
\author{Claas Willem Visser}
\affiliation{Physics of Fluids Group, Faculty of Science and Technology, MESA+ Institute, University of Twente, P.O. Box 217, 7500 AE Enschede, The Netherlands.}
\author{Henri Lhuissier}
\affiliation{Laboratoire Mati\`ere et Syst\`emes Complexes, Universit\'e Paris-Diderot / CNRS, F-75205 Paris Cedex 13, France.}
\author{Chao Sun}
\affiliation{Physics of Fluids Group, Faculty of Science and Technology, MESA+ Institute, University of Twente, P.O. Box 217, 7500 AE Enschede, The Netherlands.}
\author{Jacco H. Snoeijer}
\affiliation{Physics of Fluids Group, Faculty of Science and Technology, MESA+ Institute, University of Twente, P.O. Box 217, 7500 AE Enschede, The Netherlands.}
\affiliation{Mesoscopic Transport Phenomena, Eindhoven University of Technology, Den Dolech 2, 5612 AZ Eindhoven, The Netherlands.}
\author{Emmanuel Villermaux}
\affiliation{IRPHE, Aix-Marseille Universit\'e, 13384 Marseille Cedex 13, France.}
\affiliation{Institut Universitaire de France, 75005 Paris, France.}
\author{Detlef Lohse} \email[]{d.lohse@utwente.nl}
\author{Hanneke Gelderblom} \email[]{h.gelderblom@utwente.nl}
\affiliation{Physics of Fluids Group, Faculty of Science and Technology, MESA+ Institute, University of Twente, P.O. Box 217, 7500 AE Enschede, The Netherlands.}

\date{\today}

\begin{abstract}
We show how the deposition of laser energy induces propulsion and strong deformation of an absorbing liquid body.
Combining high speed with stroboscopic imaging, we observe that a millimeter-sized dyed water drop hit by a millijoule nanosecond laser pulse propels forward at several meters per second and deforms until it eventually fragments.
The drop motion results from the recoil momentum imparted at the drop surface by water vaporization.
We measure the propulsion speed and the time-deformation law of the drop, complemented by boundary-integral simulations.
The drop propulsion and shaping are explained in terms of the laser-pulse energy, the drop size, and the liquid properties.
These findings are, for instance, crucial for the generation of extreme ultraviolet light in nanolithography machines.
\end{abstract}

%
%
%

\maketitle

\section{Introduction}
Laser-induced phase change in liquids can lead to a violent response: deformation and disruption of the liquid body followed by the ejection of matter.
The complete vaporization or even explosion of micrometer-sized drops can result from the linear absorption of laser energy \cite{Pinnick:1990, Kafalas:1973, Kafalas:1973b}.
Self-focussing and dielectric breakdown may lead to plasma formation in transparent drops \cite{Zhang:1987, Favre:2002, Geints:2010, Lindinger:2004}.
Laser impact has been used to generate liquid motion by vaporization or plasma formation in confined geometries \cite{Vogel:1996, Sun:2009, Tagawa:2012}, sessile drops \cite{Thoroddsen:2009}, and biological matter \cite{Vogel:2003a,Apitz:2005,Horneffer:2007}.

\begin{figure}
	\includegraphics{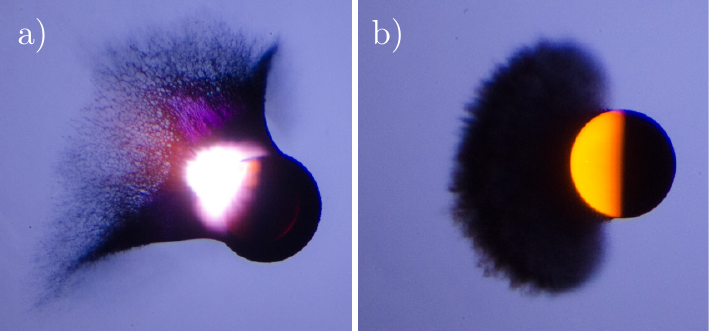}
	\caption{(Color online) Laser pulses ($\lambda = 532\,\mathrm{nm}$) impacting from the left on magenta-dyed water drops of radius $R_0 = 0.9\,\mathrm{mm}$.
	Images are taken $30\,\upmu\mathrm{s}$ after impact with a color camera and diffusive backlight illumination.
	(a) White plasma glow \cite{Kennedy:1997} and violent ablation from the drop induced by a focused laser beam.
	(b) Fluorescence of the dye and ablation at the drop surface due to local boiling induced by a uniform laser irradiation.}
	\label{fig:0} 
\end{figure}

Here, we show how the absorption of laser energy by an unconfined liquid drop induces a rapid phase change (see Fig.~\ref{fig:0}), which in turn controls the propulsion, expansion, and fragmentation of the drop.	
A key application of the drop shaping by laser impact is found in laser-produced plasma light sources for extreme ultraviolet (EUV) nanolithography \cite{Mizoguchi:2010, Banine:2011}.
In these sources the shape, position, and stability of a liquid tin body directly affect the conversion efficiency of liquid tin to a plasma that emits EUV light.

The detailed understanding of the hydrodynamic response of an opaque liquid drop to laser impact poses two fundamental challenges. 
First, one needs to resolve how momentum is transferred from the laser to the drop.
Second, the subsequent deformation dynamics and fragmentation of the drop after impact have to be quantified.
Although drop impact onto a solid substrate has been studied thoroughly (for a selection, see e.g. \cite{Clanet:2004, Yarin:2006, Villermaux:2011, Tsai:2011, Visser:2014, Tran:2013, Riboux:2014}), no consensus on the deformation dynamics has yet been reached and only few studies \cite{Yarin:2006, Xu:2007, Villermaux:2007, Villermaux:2009, Villermaux:2011} focused on the fragmentation.
%
%

\section{Experimental methods}
\begin{figure}[!ht]
	\includegraphics{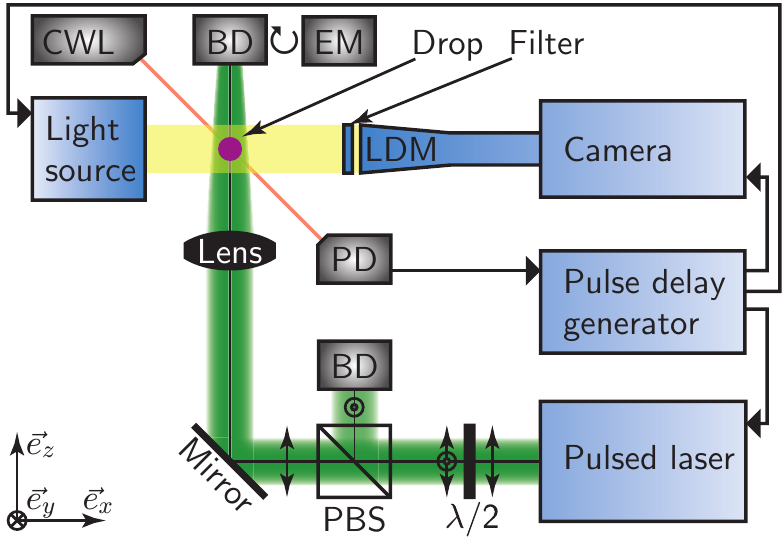}
	\caption{(Color online) Sketch of the experimental setup in top view.
	The \textbf{drop} (black/magenta ink, IJC-5900/5920 by Sensient) is generated with a capillary tube connected to a syringe pump (PHD2000 by Harvard Apparatus, not shown).
	A continuous-wave laser (\textbf{CWL}, CPS196 by Thorlabs) and a photo diode (\textbf{PD}, PDA36A by Thorlabs) serve as light barrier to trigger on the falling drop.
	The lab equipment is synchronized by a high-precision \textbf{pulse delay generator} (BNC575 by Berkeley Nucleonics) according to the indicated signal path.
	The \textbf{pulsed laser} is a frequency-doubled Nd:YAG laser (Evergreen 140 by Quantel) with a pulse duration $\tau_\mathrm{p} = 10\,\mathrm{ns}$ emitting at a wavelength $\lambda=532\,\mathrm{nm}$.
	Attenuation of the laser energy is accomplished by a zero-order half-wave plate ($\boldsymbol{\lambda}\mathbf{/2}$), a polarizing beam splitter (\textbf{PBS}), and a beam dump (\textbf{BD}).
	The laser-pulse energy is measured by an energy meter (\textbf{EM}, QE12 by gentec-eo).
	The circle and arrow symbols along the optical axis respectively indicate the S- and P-component of the linearly-polarized laser beam that is focused by a plano-convex \textbf{lens} with a focal length of $f=125\,\mathrm{mm}$.
	Side-view images ($yz$-plane) are taken with a long-distance microscope (\textbf{LDM}, 12x Zoom by Navitar), a high-speed \textbf{camera} (FASTCAM SA-X2 by Photron), and a continuous \textbf{light source} (LS-M352A by Sumita).
	Stroboscopic images are acquired by a CCD \textbf{camera} (PCO1300 by PCO AG) combined with a \textbf{light source} (NANOLITE KL-K by HSPS) that delivers a high-intensity light pulse of $8\,\mathrm{ns}$.
	A notch \textbf{filter} protects the imaging equipment from scattered laser light.
	\label{fig:ExperimentalSetup}} 
\end{figure}

Our model system consists of a highly-absorbing drop that is hit by a pulsed laser beam.
In Fig.~\ref{fig:ExperimentalSetup} an overview of the experimental setup is shown.
The drop detaches from a capillary, falls, and relaxes to a spherical shape with radius $R_0=0.9\,\mathrm{mm}$.
While it falls down, the drop masks a photodiode that then generates a reference trigger for the pulsed laser, cameras, and light source.
The $\vec{e}_z$-axis of the laser beam is aligned orthogonally to the $\vec{e}_y$-axis defined by the falling drop and the $\vec{e}_x$-axis of the imaging optics.
The $xz$-plane in which the laser beam propagates is below the plane of the trigger laser and the pinch-off point at the capillary tube.

The drop consists of dyed water with a density $\rho = \mathrm{998\,kg/m^3}$ and surface tension $\gamma = \mathrm{72\,mN/m}$ assumed to be equal to the properties of pure water. The typical penetration depth of the laser light into the dyed drop is $\delta\sim 10\,\mathrm{\upmu m}\ll R_0$ \cite{Tagawa:2012}, which ensures that the laser energy is absorbed in a thin layer close to the drop surface.
The laser-pulse energy is varied between $0$ and $120\,\mathrm{mJ}$ by an optical attenuator based on a half-wave plate and a polarizing beam-splitter.
The relation between the laser-pulse energy at the drop location and the settings of the attenuator is determined in separate measurements, for which the top beam dump shown in Fig.~\ref{fig:ExperimentalSetup} is replaced by an energy meter.
A focusing lens decreases the beam diameter to twice the drop size in order to achieve a uniform but high-intensity illumination of the drop.
To ensure the drop is placed at the center of the laser beam, the drop position is optimized such that the drop-shape evolution is axi-symmetric with respect to $\vec{e}_z$ and the propulsion speed is maximum.

The energy $E$ that is actually absorbed by the drop is computed from a beam-profile measurement and ray-tracing \cite{sup}.
The typical beam fluence $1\,\mathrm{J/cm}^2$ is well below the dielectric breakdown and self-focusing thresholds reported for water with focussed nanosecond laser pulses \cite{Kennedy:1997, Vogel:1996}.
Consistently, we observe a plasma only when the laser beam is tightly focused inside the drop (Fig.~\ref{fig:0}a, see also \cite{Vogel:2003a, Apitz:2005}), but not for a uniform irradiation (Fig.~\ref{fig:0}b).
To visualize the wavelengths in the visible spectrum that are emitted by the drop shown in Fig.~\ref{fig:0} we use a magenta-colored ink as a dye and a color camera. For all experiments leading to quantitative results we use a black-colored ink to suppress fluorescence.

The post-impact dynamics of the drop (Fig.\ \ref{fig:1}a) is observed from a side view ($\vec{e}_z$-axis in Fig.~\ref{fig:ExperimentalSetup}) with a long-distance microscope, a high-speed camera operated at a frame rate of $20\,000$ frames per second and a continuous light source.
Detailed information in the first microseconds after impact is obtained by operating a camera in stroboscopic mode with a flash lamp that delivers a high-intensity light pulse of $8\,\mathrm{ns}$ (Fig.\ \ref{fig:1}b).
We record stroboscopic videos by performing a single impact experiment per video frame while changing the time delay between the laser impact and the pulsed light source.
For both cameras used the size of the field of view is $16\,\mathrm{x}\,10\,\mathrm{mm}^2$, which yields a pixel resolution of $16\,\upmu\mathrm{m}$ per pixel.
%
%

\section{Results \& interpretation}
\begin{figure*}
	\includegraphics{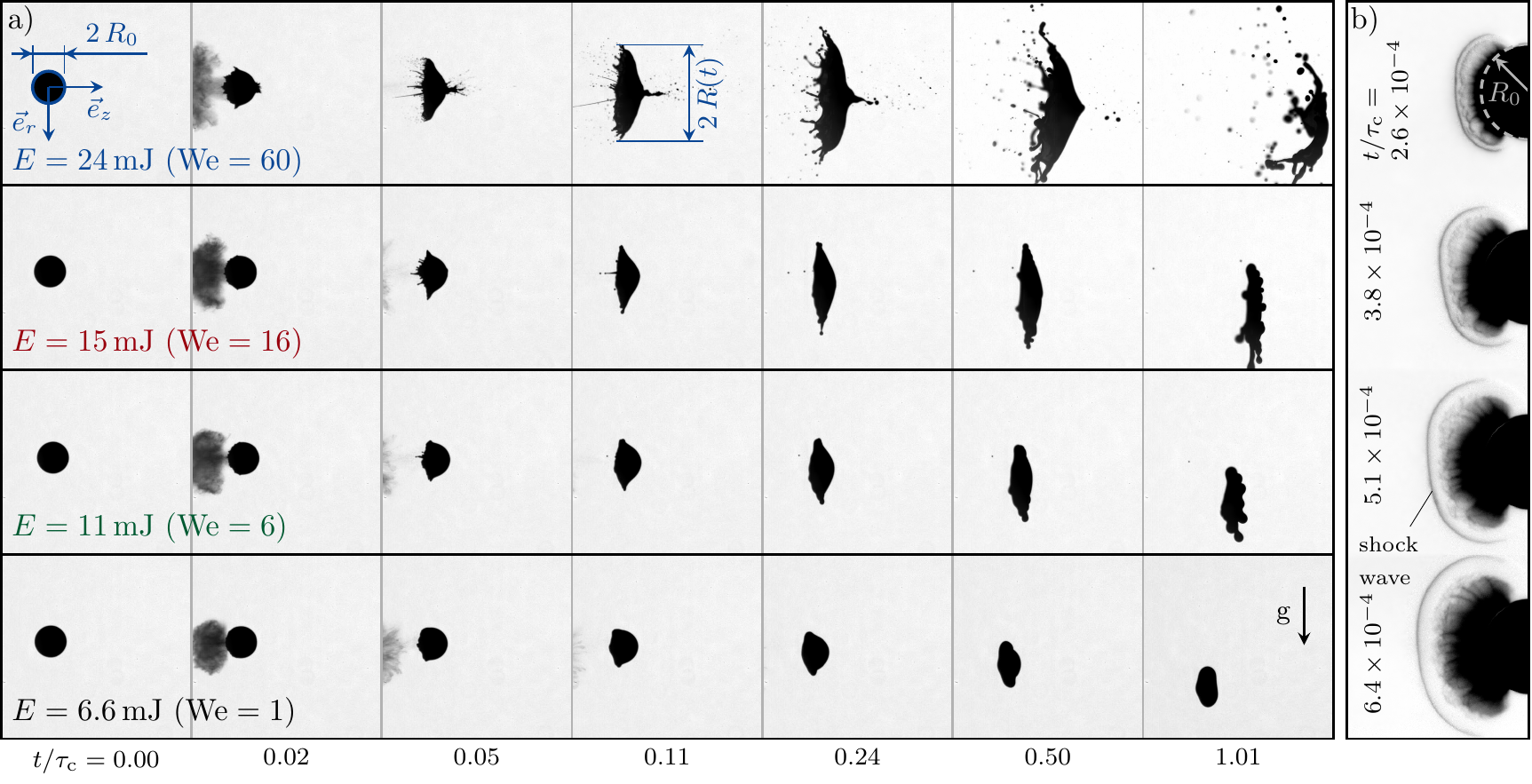}
	\caption{(Color online) Side-view of a dyed water drop with initial radius $R_0 = 0.9\,\mathrm{mm}$ hit at $t = 0$ by a laser pulse propagating from left to right ($\vec{e}_z$).
	(a) Drop shape dynamics for pulse energies increasing from bottom to top. $E$ is the energy that is actually absorbed by the drop, $\mathrm{We}$ is the Weber number of the propelled drop (see text).
	The images are taken at a frame rate of 20.000 frames per second ($\tau_\mathrm{c} = $3.5\,ms).
	As the laser ablates the front of the drop a mist cloud is ejected backward (-$\vec{e}_z$) while the remainder of the drop is propelled forward ($\vec{e}_z$) and expands radially ($\vec{e}_r$).
	For small $E$ the drop retracts after the initial expansion and no break-up occurs.
	For $E=24\,\mathrm{mJ}$ the edge destabilizes before it retracts and the drop fragments.
	(b) Close-up view of the drop surface for $E = 24\,\mathrm{mJ}$ revealing the shock wave in the air and the mist cloud development at early times (increasing from top to bottom).}
	\label{fig:1} 
\end{figure*}

The drop dynamics for different pulse energies is shown in Fig.~\ref{fig:1}. 
On impact, the surface of the drop hit by the laser emits a shock wave into the air (Fig.~\ref{fig:1}b).
The shock wave is followed by the ejection of a mist cloud of small drops that is visible as a gray-to-black haze in the images and persists for several microseconds.
Subsequently, the mist is expelled while the drop propels in the opposite direction (Fig.~\ref{fig:1}a).
At the same time the drop flattens and expands in the radial direction before it either retracts, for low pulse energy, or fragments, for large energy. 
%
%
		
\begin{figure}
	\includegraphics{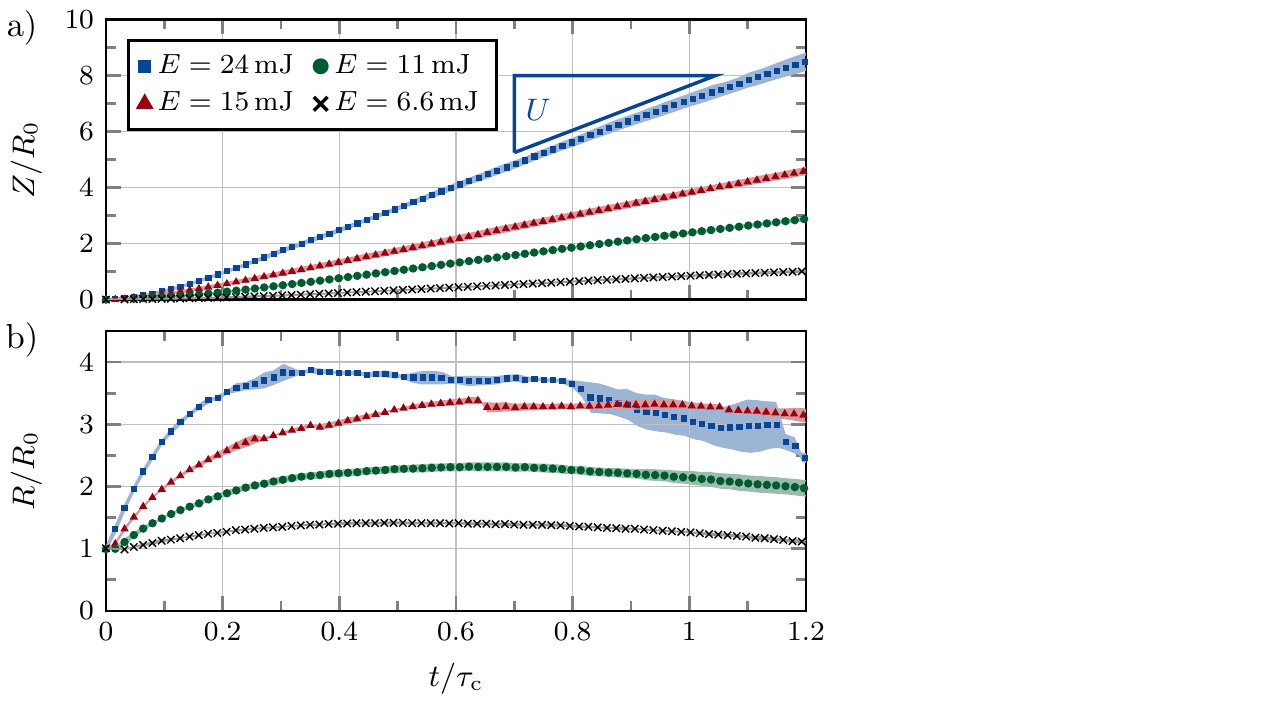}
	\caption{(Color online) Center-of-mass displacement~$Z$ (along $\vec{e}_z$, see Fig.~\ref{fig:1}) (a) and radial expansion~$R$ (b) as a function of time for different absorbed energies.
	The corresponding image sequences are shown in Fig.~\ref{fig:1}.
	Each point is averaged over two experiments and the shaded area indicates the difference between the two.
	The apparent acceleration in $Z$ for $t/\tau_\mathrm{c} < 0.2$ is an artifact of the method used to determine the center-of-mass position \cite{sup}.
	The large deviation in $R$ for $E=24$\,mJ illustrates the statistical nature of the fragmentation.
	For $E=15$\,mJ events of drop ejection from the edge are visible at $t/\tau_\mathrm{c} = 0.24$, $0.36$ and $0.63$.}
	\label{fig:2} 
\end{figure}

We quantify the drop motion by measuring the displacement $Z(t)$ of the drop center-of-mass and the drop radius $R(t)$ (defined in Fig.~\ref{fig:1}a) for the first milliseconds after impact.
As Fig.~\ref{fig:2}a shows, the drop is propelled at a constant speed $U$ that increases with increasing pulse energy up to $2.0\,{\rm m/s}$.
The accompanying deformation of the drop occurs on the inertial time-scale $\tau_{\rm i} = R_0/U \sim 10^{-4}$ to $10^{-3}\,\mathrm{s}$ (Fig.~\ref{fig:2}b) and is eventually slowed down by surface tension on the capillary time-scale $\tau_\mathrm{c}=\sqrt{\rho R_0^3/\gamma} =3.5\,\mathrm{ms}$.
Both the initial deformation rate $\tau_{\rm i}^{-1}$ and the maximal extension $R_{\rm max}$ increase with increasing pulse energy.
We emphasize the clear separation of time-scales
\begin{equation}
	\tau_{\rm p} \ll \tau_{\rm e}  \ll \tau_{\rm i} < \tau_{\rm c} 
	\label{eq:times}
\end{equation}
between the successive steps, namely, the laser pulse, the ejection of matter (on time scale $\tau_{\rm e} \sim 10^{-5}\,\mathrm{s}$), the initial deformation of the drop, and its capillary retraction.  
%
%

To explain the relation between the drop propulsion speed, the radial expansion, and the laser energy one needs to understand the mechanism that propels the drop.
Surely, both the optical radiation pressure from the laser and the thermal radiation pressure caused by the heating of the drop surface are insignificant \cite{radscale, Delville:2009}.
The motion actually results from the recoil due to the partial vaporization of the drop: since the highly-absorbent dye ensures that the laser energy is absorbed in a superficial layer on one side of the drop, the vapor expulsion is mainly unidirectional and consequently transfers momentum to the remainder of the drop.

The light energy is absorbed by a liquid mass $\sim \rho R_0^2 \delta$ set by the penetration depth of the laser.
On the time scale $\tau_{\rm e}$, both diffusive and radiative heat transfers are negligible (the thermal diffusion length is much smaller than $\delta$ \cite{thermscale}).
Since the beam profile is flat, and neither the focusing due to the drop interface curvature nor nonlinear optical effects (self-focusing or electric breakdown) are significant, we consider the energy deposition in the superficial layer to be close to uniform.
This energy is sufficient to heat the liquid from the ambient temperature $T_0 = 293$\,K to the boiling temperature $T_{\rm b} \simeq 393\,$K, but not to vaporize all of it: only a certain fraction $\beta$ actually vaporizes.
The energy balance therefore reads $E\sim\rho R_0^2\delta [c_\mathrm{v} (T_\mathrm{b} - T_0)+ \beta \Delta H]$, where $c_\mathrm{v} = 4.0\,{\rm kJ/(kg\,K)}$ and $\Delta H = 2.25\,{\rm MJ/kg}$ are, respectively, the specific heat capacity and latent heat of vaporization of the liquid.

\begin{figure}
	\includegraphics{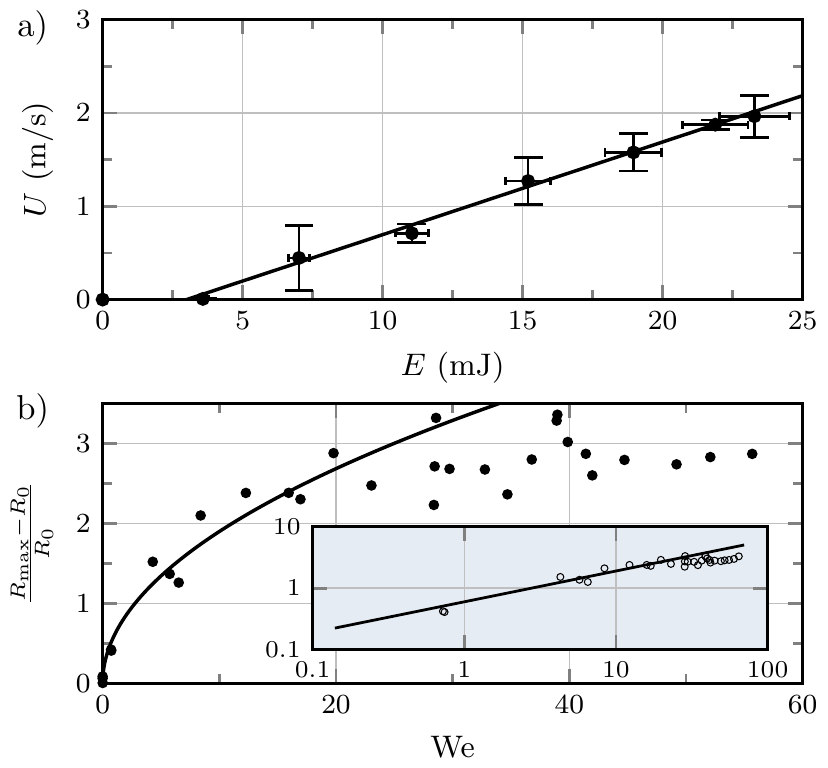}
 	\caption{(Color online) (a) Propulsion speed of the drop as a function of the absorbed laser energy.
	Each point represents at least four experiments, the error bars indicate the standard deviation.
	The solid line is equation (\ref{eq:1}) with a prefactor of $0.4$.
	(b) Maximal relative expansion $R_\mathrm{max}/R_0-1$ for individual experiments as a function of the Weber number in linear and logarithmic (inset) scales.
	The solid line is equation (\ref{eq:2}) with a prefactor of $0.6$.
	For large Weber numbers a saturation is observed due to the fragmentation of the sheet.}
	\label{fig:3} 
\end{figure}

In all our experiments a mist cloud is observed, which is a clear signal of a local boiling of the drop. We therefore assume that to get propulsion, a threshold energy $E_\mathrm{th} \sim \rho R_0^2 \delta c_\mathrm{v}\, (T_\mathrm{b}-T_0)\approx 3\,\mathrm{mJ}$ has to be absorbed by the superficial layer to heat the liquid to the boiling point, which is in good agreement with the threshold for propulsion observed in our experiments (Fig.~\ref{fig:3}a).
Any additional energy deposited in the superficial layer is used to vaporize a mass of liquid $m\sim \beta\rho R_0^2 \delta \sim (E-E_\mathrm{th})/\Delta H$.
An upper limit for the proposed scaling is given by $E/E_\mathrm{th} \sim 1+\Delta H/[c_\mathrm{v}(T_\mathrm{b} - T_0)]\approx 8$, in which case the absorbed energy is sufficient to evaporate the entire heated liquid layer (i.e.~$\beta=1$).
Any increase in $E$ beyond this point would lead to a superheated or even a critical phase, in which case the opaque mist cloud would not be observed \cite{Apitz:2005}.

For $0<\beta<1$, which is the case of our experiments, the remaining part of the heated layer that is not vaporized is expelled as a mist of small drops.
We assume that the liquid vaporizes at $T_{\rm b}$ and that the vapor is expelled at the thermal speed $u = \sqrt{k_b T_\mathrm{b}/\mu}\approx 400\,\mathrm{m/s}$, where $k_b\simeq1.38\times10^{-23}\,$J/K is the Boltzmann constant and $\mu =2.99\times 10^{-26}\,$kg is the molecular mass of water.
This expelled vapor propels the remainder of the drop. Momentum conservation $mu = \rho R_0^3 U$ yields
\begin{equation}
	U\sim \frac{E-E_\mathrm{th}}{\rho\, R_0^3\, \Delta H}\,u\label{eq:1},
\end{equation}
that is, an increase in $U$ proportional to that in $E$.
Figure \ref{fig:3}a shows that this scaling argument, with a prefactor of $0.4$, is in good agreement with our experimental data.
			
With a description of the propulsion at hand, we now turn to the drop deformation.
The expansion dynamics is directly affected by surface tension, which promotes the retraction and possibly the fragmentation of the drop.
The key parameter describing the expansion is therefore the Weber number of the motion induced by the laser $\mathrm{We}=\rho R_0 U^2/\gamma$, which compares the drop displacement kinetic energy to its surface energy.
In our experiments $1\leq\mathrm{We}\leq60$. The impulsive acceleration of our drop from 0 to $U$ is similar to the impulsive stop of a drop impacting a solid with velocity $U$.
We therefore use the momentum-based scaling derived by \cite{Clanet:2004, Villermaux:2011} for drop impact on solids to express the maximal radial expansion
\begin{equation}
	\frac{R_\mathrm{max}-R_0}{R_0}\sim \mathrm{We}^{1/2} \sim \sqrt{\frac{\rho R_0 u^2}{\gamma}}\frac{E-E_\mathrm{th}}{\rho\, R_0^3\, \Delta H},\label{eq:2}
\end{equation}
in which the expression in terms of $E$ directly comes from (2).
Expression (3), with a prefactor of $0.6$, is in good agreement with our experimental data up to $\mathrm{We}\sim 40$, when the drop starts fragmenting and the maximum expansion saturates (see Fig.~\ref{fig:3}b).
The scaling (3) has already been observed for drop impact onto solid substrates with negligible friction \cite{Villermaux:2011}.
The present setup is however fundamentally different since, as mentioned above, the typical impact timescale, during which the drop accelerates, is decoupled from the inertial timescale: $\tau_{\rm e} \ll \tau_{\rm i}$.
%
%

\section{Numerical results}
To confirm that the interaction of the laser pulse with the drop can be modeled as a short recoil-pressure pulse exerted on the drop surface, we perform boundary integral (BI) simulations \cite{Bergmann:2009, Gekle:2010, Bouwhuis:2012}.
We assume that the flow inside the drop is inviscid, irrotational, and incompressible, and solve the resulting Laplace equation for the flow potential.
The method assumes axi-symmetry and therefore cannot be used to study the eventual fragmentation of the drop, but it does capture the initial phase of the drop deformation.

The laser pulse is modeled by applying a pressure boundary condition at the drop surface for a time duration $\tau_\mathrm{e}\ll \tau_{\rm i}$.
We use a Gaussian pressure profile with a length-scale based on the measured laser-beam profile and a pressure scale set to match the propulsion velocity observed in the experiment.
This pressure scale is prescribed by the momentum conservation $p R_0^2 \tau_\mathrm{e} \sim \rho R_0^3 U$ (the prefactor can be obtained analytically \cite{Gelderblom:2015}). From (\ref{eq:1}) this recoil pressure can readily be expressed in terms of the absorbed energy.

\begin{figure}
	\includegraphics{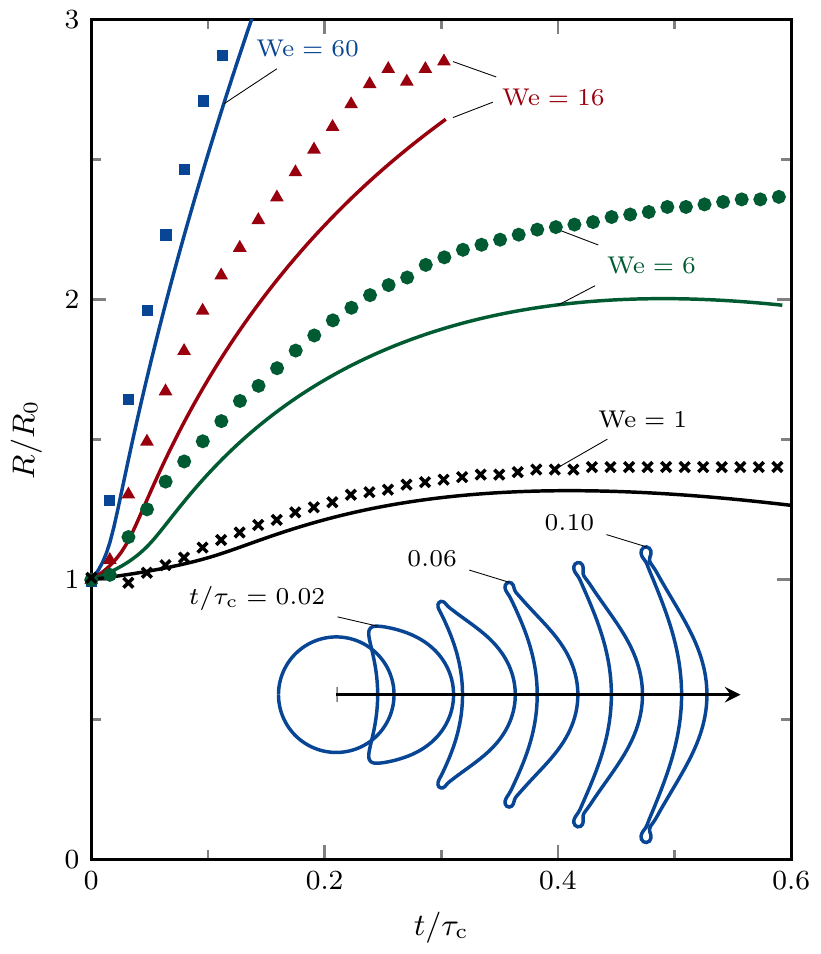}
 	\caption{(Color online) Radial expansion of the drop: experiments (markers) and BI simulations (solid lines).
	The corresponding image sequences are shown in Fig.~\ref{fig:1}.
	The inset shows the drop shape evolution from BI for $\mathrm{We}=60$ with an exaggerated center-of-mass displacement to separate the successive shapes.
	The simulations are stopped when the local sheet thickness becomes too thin to resolve the dynamics accurately.}
	\label{fig:4} 
\end{figure}

The numerical drop shape evolution is shown in Fig.~\ref{fig:4}.
It illustrates the added value of the simulations: not only the two-dimensional projection of the drop shape, but also the spatial and temporal evolution of the sheet thickness can be extracted, which is crucial when it comes to study fragmentation \cite{Villermaux:2011}.
Moreover, Fig.~\ref{fig:4} shows that the BI model quantitatively predicts the radial drop expansion observed for different Weber numbers.
This confirms that a pressure pulse applied at the drop surface for a time much shorter than the hydrodynamic time scales ($\tau_{\rm i}$ and $\tau_{\rm c}$) is indeed sufficient to describe the hydrodynamic response of a drop to the impact of a laser pulse.
%
%

\section{Conclusions}
We have shown that an opaque free-falling drop hit by a laser pulse propels and expands until fragmentation occurs.
In the present case the laser energy is absorbed in a superficial layer of the drop such that the deposited energy per unit mass $E/\rho R_0^2\delta \sim 0.1$ to $1\,\rm{MJ/kg}$ is comparable to the specific latent heat of vaporization.
As a consequence, drop motion is induced by the recoil due to vaporization on the face of the drop that is hit by the laser.
This results in a propulsion speed and a maximal radius of expansion that are both proportional to the pulse energy.
The expansion dynamics is limited by surface tension and is similar to that of a drop impacting a solid, although with a laser pulse momentum transfer takes place on a much shorter time scale.
Laser-induced drop fragmentation and the influence of the beam focussing require detailed studies and are left for future work \cite{Gelderblom:2015}. 
All results reported here should transpose directly to the shaping of liquid tin drops in EUV light sources.  
In a regime where a plasma is generated the propulsion mechanism may change, however, the Weber number remains the key parameter governing the hydrodynamic response.

\begin{acknowledgments}
We thank Chris Lee, Andrea Prosperetti, and Guillaume Lajoinie for fruitful discussions.
This work is part of an Industrial Partnership Programme of the Foundation for Fundamental Research on Matter (FOM), which is financially supported by the Netherlands Organization for Scientific Research (NWO).
This research programme is co-financed by ASML. W.B.~and J.H.S.~acknowledge support from NWO through VIDI Grant No.~11304.
\end{acknowledgments}

\bibliography{main}

\begin{thebibliography}{35}
\bibitem [{the()}]{thermscale}%
  \BibitemOpen
  \href@noop {} {\bibinfo  {journal} {On the time scale of vapor ejection
  $\tau_\mathrm{e}$ heat typically diffuses over a length $\sqrt{\kappa
  \tau_\mathrm{e}}\sim 10^{-6}\,\mathrm{m} \ll \delta$, where $\kappa$ is the
  thermal diffusivity of water}\ }\BibitemShut {NoStop}%
\end{thebibliography}

\end{document}